\documentclass[conference]{IEEEtran}
\usepackage[T1]{fontenc}
\usepackage[utf8]{inputenc}
\usepackage{amssymb}
\usepackage[pdftex]{graphicx}
\usepackage{subcaption} 
\usepackage{amsmath}
\usepackage{amsthm}
\usepackage{mathtools}
\usepackage{array}
\usepackage[noadjust]{cite}
\usepackage{algorithm}
\usepackage{algpseudocode}
\usepackage{xcolor}
\usepackage{tikz}
\usepackage{pgfplots}
\usepackage[acronym]{glossaries}
\usepackage{dutchcal}
\pgfdeclarelayer{background}
\pgfdeclarelayer{foreground}
\pgfsetlayers{background,main,foreground}
\usepgfplotslibrary{fillbetween}
\usepgfplotslibrary{external}
\usetikzlibrary{spy,backgrounds}
\usepackage[binary-units=true]{siunitx}
\usepackage{balance}
\makeatletter
\let\MYcaption\@makecaption
\makeatother

\usepackage{subcaption}

\makeatletter
\let\@makecaption\MYcaption
\makeatother
\algnewcommand{\LineComment}[1]{\(\triangleright\) #1}

\DeclareSIUnit\frame{frame}
\DeclareSIUnit\PRB{PRB}

\newacronym{SNR}{SNR}{signal-to-noise ratio}
\newacronym{LDPC}{LDPC}{low-density parity-check}
\newacronym{QAM}{QAM}{quadrature amplitude modulation}
\newacronym{QPSK}{QPSK}{quadrature phase-shift keying}
\newacronym{BICM}{BICM}{bit-interleaved coded modulation}
\newacronym{BCE}{BCE}{binary cross-entropy}
\newacronym{LLR}{LLR}{log-likelihood ratio}
\newacronym{BMI}{BMI}{bit-wise mutual information}
\newacronym{KL}{KL}{Kullback–Leibler}
\newacronym{BMD}{BMD}{bit-metric decoding}
\newacronym{BER}{BER}{bit error rate}
\newacronym{NN}{NN}{neural network}
\newacronym{GS}{GS}{geometric shaping}
\newacronym{SIP}{SIP}{superimposed pilot}
\newacronym{iid}{i.i.d.\@}{independent and identically distributed}
\newacronym{SGD}{SGD}{stochastic gradient descent}
\newacronym{wrt}{w.r.t.\@}{with respect to}
\newacronym{MAP}{MAP}{maximum a posteriori}
\newacronym{LMMSE}{LMMSE}{linear minimum mean square error}
\newacronym{AWGN}{AWGN}{additive white Gaussian noise}
\newacronym{RBF}{RBF}{Rayleigh block fading}
\newacronym{OFDM}{OFDM}{orthogonal frequency division multiplexing}
\newacronym{3GPP}{3GPP}{3rd Generation Partnership Project}
\newacronym{5GNR}{5G NR}{5G New Radio}
\newacronym{PRB}{PRB}{physical resource block}
\newacronym{CSI}{CSI}{channel state information}
\newacronym{MSE}{MSE}{mean squared error}
\newacronym{DC}{DC}{direct current}
\newacronym{PAPR}{PAPR}{peak-to-average power ratio}
\newacronym{DMRS}{DMRS}{demodulation reference signal}
\newacronym{TTI}{TTI}{transmission time interval}
\newacronym{CDF}{CDF}{cumulative distribution function}
\newacronym{RE}{RE}{resource element}
\newacronym{MIMO}{MIMO}{multiple-input multiple-output}
\newacronym{IDFT}{IDFT}{inverse discrete Fourier transform}
\newacronym{DFT}{DFT}{discrete Fourier transform}
\newacronym{CP}{CP}{cyclic prefix}
\newacronym{FFT}{FFT}{fast Fourier transform}
\newacronym{ISI}{ISI}{inter-symbol interference}
\newacronym{ICI}{ICI}{inter-carrier interference}
\newacronym{LOS}{LoS}{line-of-sight}
\newacronym{NLOS}{NLoS}{non-line-of-sight}
\newacronym{ACLR}{ACLR}{adjacent channel leakage ratio}
\newacronym{RRC}{RRC}{root-raised-cosine}
\newacronym{PED}{PED}{power envelope distribution}
\newacronym{ML}{ML}{machine learning}
\newacronym{CCDF}{CCDF}{complementary cumulative distribution function}
\newacronym{NPD}{NPD}{normalized power dispersion}
\newacronym{PSD}{PSD}{power spectral density}
\renewcommand{\vec}[1]{\mathbf{#1}}
\newcommand{\vecs}[1]{\boldsymbol{#1}}

\newcommand{\rv}{\vec{r}}
\newcommand{\sv}{\vec{s}}

\newcommand{\xv}{\vec{x}}

\newcommand{\gammav}{\vecs{\gamma}}

\newcommand{\thetav}{\vecs{\theta}}
\newcommand{\psiv}{\vecs{\psi}}

\newcommand{\Am}{\vec{A}}
\newcommand{\Bm}{\vec{B}}

\newcommand{\Em}{\vec{E}}

\newcommand{\Xm}{\vec{X}}

\newcommand{\Cc}{{\cal C}}

\newcommand{\Lc}{{\cal L}}

\newcommand{\Sc}{{\cal S}}

\newcommand{\Uc}{{\cal U}}

\newcommand{\CC}{\mathbb{C}}

\newcommand{\RR}{\mathbb{R}}

\newcommand{\ZZ}{\mathbb{Z}}

\newcommand{\htp}{^{\mathsf{H}}}
\newcommand{\tp}{^{\mathsf{T}}}

\newcommand{\LB}{\left(}
\newcommand{\RB}{\right)}
\newcommand{\LP}{\left\{}
\newcommand{\RP}{\right\}}
\newcommand{\LSB}{\left[}
\newcommand{\RSB}{\right]}

\renewcommand{\ln}[1]{\mathop{\mathrm{ln}}\LB #1\RB}
\renewcommand{\log}[1]{\mathop{\mathrm{log_2}}\LB #1\RB}
\renewcommand{\exp}[1]{\mathop{\mathrm{exp}}\LB #1\RB}

\newcommand{\EE}{{\mathbb{E}}}

\newcommand\abs[1]{\left| #1 \right|}

\begin{document}
\title{End-to-end Waveform Learning Through Joint Optimization of Pulse and Constellation Shaping}

\IEEEoverridecommandlockouts 

\author{\IEEEauthorblockN{Fay\c{c}al Ait Aoudia\IEEEauthorrefmark{1} and Jakob Hoydis\IEEEauthorrefmark{3}\thanks{Work carried out while J. Hoydis was with Nokia Bell Labs.}}
\IEEEauthorblockA{\IEEEauthorrefmark{1}Nokia Bell Labs, Paris-Saclay, 91620 Nozay, France\\
\IEEEauthorrefmark{3}NVIDIA, 06906 Sophia Antipolis, France\\
 faycal.ait\_aoudia@nokia-bell-labs.com, jhoydis@nvidia.com}
}

\maketitle

\begin{abstract} As communication systems are foreseen to enable new services such as joint communication and sensing and utilize parts of the sub-THz spectrum, the design of novel waveforms that can support these emerging applications becomes increasingly challenging.
We present in this work an end-to-end learning approach to design waveforms through joint learning of pulse shaping and constellation geometry, together with a \gls{NN}-based receiver.
Optimization is performed to maximize an achievable information rate, while satisfying constraints on out-of-band emission and power envelope.
Our results show that the proposed approach enables up to orders of magnitude smaller \glspl{ACLR} with \glspl{PAPR} competitive with traditional filters, without significant loss of information rate on an \gls{AWGN} channel, and no additional complexity at the transmitter.

\end{abstract}
\glsresetall

\section{Introduction}

While 5G communication infrastructures are actively deployed, academic and industrial research has shifted focus towards the next generation of cellular communication systems~\cite{9040431}.
To support the numerous services envisioned for future systems, such as sub-THz communication or joint communication and sensing/power transfer, the design of new waveforms is required.
As an example, communication in the sub-THz bands involves lower power amplifier efficiency, higher phase noise, and meeting strict regulation on out-of-band emission.
While most of current communication systems rely on \gls{OFDM} due to its very efficient implementation, its high \gls{PAPR} and \gls{ACLR} make it unsuitable for many emerging services.

We propose in this work to use end-to-end learning to design new waveforms that maximize an achievable information rate, while strictly satisfying constraints on the \gls{ACLR} and \gls{PED}.
End-to-end learning consists in implementing the transmitter, channel, and receiver as a \gls{NN}, that is trained to achieve the highest possible information rate~\cite{9118963}.
This idea has been successfully applied to many fields, including optical fiber~\cite{Karanov:18}, coding~\cite{9252949}, and more recently to achieve pilot- and \gls{CP}-less communication in \gls{OFDM} systems~\cite{aoudia2021trimming}.
In this work, a waveform is learned on the transmitter side through optimization of the constellation geometry and transmit filter, that replace, e.g., conventional \gls{QAM} and \gls{RRC} filtering.
On the receiver side, a receive filter and \gls{NN}-based detector are jointly optimized for the constellation and transmit filter.
The \gls{NN}-based detector computes \glspl{LLR} on the transmitted bits directly from the received samples.
Optimization of the constellation geometry and transmit filter are performed with constraints on the \gls{ACLR} and \gls{PED}.
To the best of our knowledge, this is the first work that demonstrates how the transmit and receive filters of a communication system can be jointly optimized through end-to-end learning to design waveforms.

We have considered an \gls{AWGN} channel and conventional \gls{RRC} filtering with Blackman windowing for benchmarking.
However, applying the proposed method to other channel models is straightforward.
The \gls{RRC} filter allows control of the tradeoff between the excess bandwidth and the magnitude of the ripples through the roll-off factor parameter, which also impacts the \gls{PED}.
Our results show that the proposed approach allows fine-grained control of the \gls{ACLR} and \gls{PED} variance, while maximizing the information rate.
Compared to the baseline, the learned waveforms enable up to three orders of magnitude lower \glspl{ACLR}, with similar \glspl{PED} and competitive information rate.

\paragraph*{Notations}
$()^*$ is the complex conjugate operator.
$\ln{\cdot}$ denotes the natural logarithm and $\log{\cdot}$ the binary logarithm.
The $(i,k)$ element of a matrix $\Xm$ is denoted by $X_{i,k}$. The $k^{th}$ element of a vector $\xv$ is $x_k$ .
The operators $()\htp$ and $()\tp$ denote the Hermitian transpose and transpose, respectively. 
Finally, $j$ denotes the imaginary unit, i.e., $j^2 = -1$.

\section{System model and baseline}
\label{sec:sys}


\subsection{System model}
\label{sec:sys_mod}

\begin{figure}
 	\centering
	\includegraphics[scale=0.7]{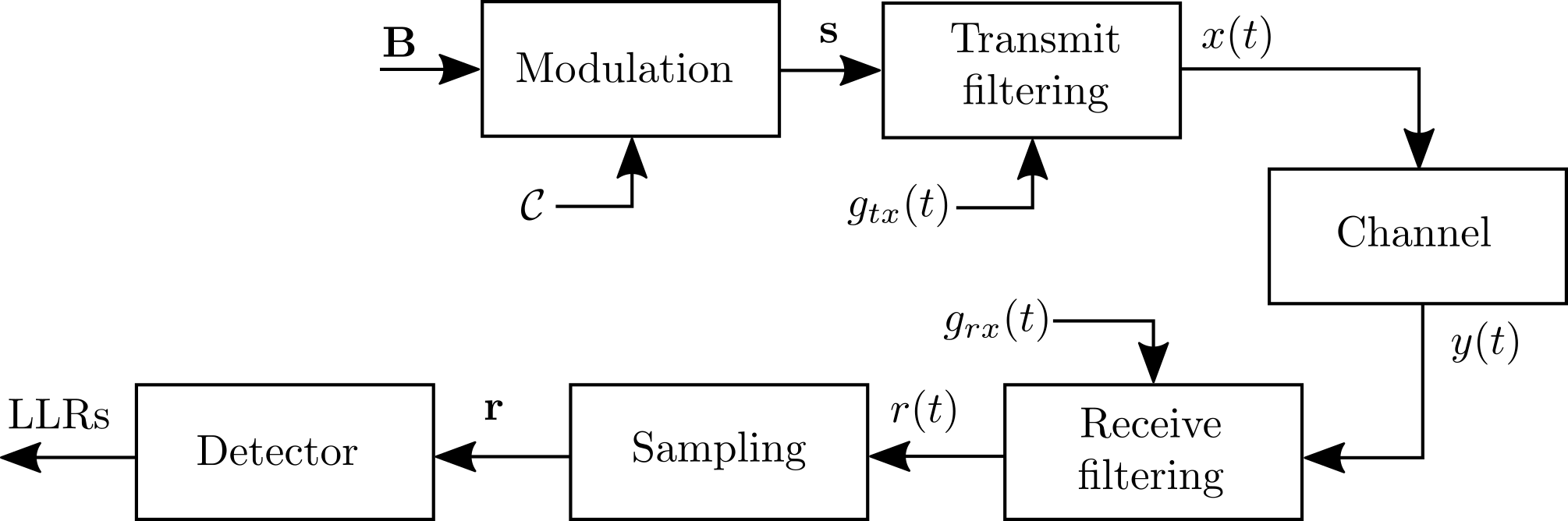}
	\caption{Single-carrier communication system, where $\Cc$ is the constellation used for modulation, $g_{tx}(t)$ the transmit filter, and $g_{rx}(t)$ the receive filter.\label{fig:sys}}
\end{figure}

A single-carrier system is considered, and the matrix of bits to be transmitted is denoted by $\Bm \in \{0,1\}^{N \times K}$, where $N$ is the number of baseband symbols forming a block, and $K$ the number of bits per channel use.
As illustrated in Fig.~\ref{fig:sys}, $\Bm$ is modulated onto a vector of baseband symbols $\sv \in \Cc^N$ according to some constellation $\Cc$, e.g., a $2^K$-\gls{QAM} or a learned constellation.
The modulated symbols are then shaped using a transmit filter $g_{tx} : t \in \RR \rightarrow \CC$ to generate the time-continuous baseband signal
\begin{equation}
	\label{eq:x}
	x(t) = \sum_{m=0}^{N-1} s_m g_{tx}\LB t-mT \RB
\end{equation}
where $T$ is the symbol period.

The time-continuous channel output is denoted by $y(t)$.
On the receiver side, the received signal is first filtered using a receive filter $g_{rx} : t \in \RR \rightarrow \CC$ to generate the signal
\begin{equation}
	r(t) = \int_{-\infty}^{+\infty} y(z) g_{rx}\LB t-z \RB dz
\end{equation}
which is then sampled with a period $T$ to form the vector $\rv \in \CC^N$.
Assuming an \gls{AWGN} channel, one has
\begin{equation}
	\label{eq:ch}
	r_m = r(mT) = \sum_l s_{m-l} \alpha \LB lT \RB + n_m
\end{equation}
where $n_m$ is the additive Gaussian noise, and the function $\alpha(\cdot)$ is the convolution between the transmit and receive filter, i.e.,
\begin{equation}
	\label{eq:alpha}
	\alpha(t) = \int_{-\infty}^{+\infty} g_{tx}(z) g_{rx}(t-z) dz.
\end{equation}
Moreover, the correlation of the additive noise sequence $\LP n_m \RP_{0\leq m \leq N-1}$ is controlled by the receive filter,
\begin{equation}
	\label{eq:noise_cov}
	\EE \LSB n_m n_{m+l}^* \RSB = N_0 \int_{-\infty}^{+\infty} g_{rx}(z) g_{rx}^*\LB z-lT \RB dz
\end{equation}
where $N_0$ is the noise power density.
As shown in Fig.~\ref{fig:sys}, $\rv$ is processed by a detector which computes \glspl{LLR} on the transmitted bits.
The \glspl{LLR} could then be fed to a channel decoder to reconstruct the transmitted bits.

\subsection{QAM with RRC filtering}
\label{sec:bsl}

Conventional transmit and receive filters are designed to minimize the \gls{ACLR}, \gls{PAPR}, and to avoid \gls{ISI} by satisfying the Nyquist \gls{ISI} criterion~\cite[Chapter~9.2.1]{proakis1994communication}, i.e.,
\begin{equation}
	\label{eq:nyquist_isi}
	\alpha(lT) =
	\begin{cases}
		1 & \text{ if } l = 0\\
		0 & \text{ otherwise}
	\end{cases}
\end{equation}
where $l \in \ZZ$.
The \gls{RRC} filter, denoted by $\text{rrc}_{\beta}(t)$, where $\beta \in (0,1)$ is the roll-off factor, is widely used as transmit and receive filter, and is considered in this work as a baseline.
It is defined such that its autoconvolution is the raised-cosine function, which satisfies the Nyquist \gls{ISI} criterion~\eqref{eq:nyquist_isi}.
The roll-off factor $\beta$ controls a tradeoff between the excess bandwidth and the magnitude of the ripples.
As the \gls{RRC} filter is not time-limited, it is windowed by the well-known Blackman windowing function $w(t)$ for practical use, which results in the following transmit and receive filters:
\begin{equation}
	g_{tx}(t) = g_{rx}(t) = \text{rrc}_{\beta}(t) w\left(\frac{t}{D}\right)
\end{equation}
where $D > 0$ is the filter duration.

After sampling, detection is performed assuming no \gls{ISI}, and using the usual \gls{AWGN} detector to compute \glspl{LLR}, i.e.,
\begin{equation}
	\text{LLR}(m,k) = \ln{\frac{\sum_{c\in\Cc(k,0)} \exp{-\frac{\abs{r_m-c}^2}{N_0}} }{ \sum_{c\in\Cc(k,1)} \exp{-\frac{\abs{r_m-c}^2}{N_0}} } }
\end{equation}
where $\text{LLR}(m,k)$ is the \gls{LLR} for the $k^{th}$ bit ($0 \leq k \leq K-1$) of the $m^{th}$ symbol ($0 \leq m \leq N-1$), and $\Cc(k,0)$ ($\Cc(k,1)$) is the subset of $\Cc$ which contains all constellation points with the $k^{th}$ bit label set to 0(1).

\section{Problem formulation}


In our setting, the transmitter includes a trainable transmit filter $g_{tx,\thetav}(t)$ with trainable parameters $\thetav$, and a trainable constellation $\Cc$.
The receiver includes a trainable receive filter $g_{rx,\psiv}(t)$ with trainable parameters $\psiv$, and an \gls{NN}-based detector with trainable parameters $\gammav$, which generates posterior probabilities on the transmitted bits $B_{m,k}$ denoted by $Q_{\gammav}(B_{m,k} | \rv)$, $0 \leq m \leq N-1$, $0 \leq k \leq K-1$.
The transmit and receive filters are both time-limited to the interval $\LB -\frac{D}{2}, \frac{D}{2} \RB$, i.e., for $t \notin \LB -\frac{D}{2}, \frac{D}{2} \RB$, $g_{tx,\thetav}(t) = g_{rx,\psiv}(t) = 0$.

The end-to-end system is trained to maximize the rate 
\begin{multline}
	\label{eq:R}
	R(\Cc,\thetav,\psiv,\gammav) \coloneqq \frac{1}{N} \sum_{m=0}^{N-1} \sum_{k=0}^{K-1} I(B_{m,k};\rv | \Cc,\thetav,\psiv)\\
	- \frac{1}{N} \sum_{m=0}^{N-1} \sum_{k=0}^{K-1}\EE_{\rv} \LSB \text{D}_{\text{KL}} \LB \Pr\LB B_{m,k} | \rv \RB || Q_{\gammav}\LB B_{m,k} | \rv \RB \RB \RSB
\end{multline}
which is an achievable rate for practical \gls{BICM} systems~\cite{bmi}.
In~\eqref{eq:R}, $I(B_{m,k};\rv | \Cc,\thetav,\psiv)$ is the mutual information between $B_{m,k}$ and $\rv$ conditioned on the constellation $\Cc$ and filters parameters $\thetav$ and $\psiv$, $\text{D}_{\text{KL}}(\cdot||\cdot)$ is the \gls{KL} divergence, $\Pr\LB B_{m,k} | \rv \RB$ is the \emph{true} posterior distribution on $B_{m,k}$ given $\rv$, and $Q_{\gammav}\LB B_{m,k} | \rv \RB$ is the posterior distribution on $B_{m,k}$ given $\rv$ \emph{approximated} by the \gls{NN}-based detector.
Intuitively, the first term in the right-hand side of~\eqref{eq:R} corresponds to an achievable rate assuming an optimal receiver, i.e., one that implements $\Pr\LB B_{m,k} | \rv \RB$.
The second term is the rate loss due to the use of a mismatched receiver $Q_{\gammav}\LB B_{m,k} | \rv \RB \neq \Pr\LB B_{m,k} | \rv \RB$.

The waveform of the transmitted signal $x(t)$~\eqref{eq:x} (where $g_{tx,\thetav}(t)$ is used for filtering) is enforced to meet practical constraints.
To begin with, the transmit filter and constellation are required to have unit energy, i.e.,
\begin{align}
	&\int_{-\frac{D}{2}}^{\frac{D}{2}} \abs{g_{tx,\thetav}(t)}^2 dt = 1 \label{eq:cst_en}\\
	&\EE_{c \sim \Uc\LB\Cc\RB} \LSB \abs{c}^2 \RSB = 1
\end{align}
where $\Uc\LB\Cc\RB$ is the uniform distribution on $\Cc$.
Focusing on the \gls{ACLR}, it is defined by
\begin{equation}
	\text{ACLR}(\thetav) \coloneqq \frac{E_O(\thetav)}{E_I(\thetav)} = \frac{1}{E_I(\thetav)}-1
\end{equation}
where $E_I(\thetav)$ is the in-band energy, $E_O(\thetav)$ the out-of-band energy, and the second equality comes from~\eqref{eq:cst_en} as it leads to $E_I(\thetav) + E_O(\thetav) = 1$.
As the sequence of modulated symbols $\sv$ is assumed to be \gls{iid}, the in-band energy is
\begin{equation}
	\label{eq:inb}
	E_I(\thetav) = \int_{-\frac{W}{2}}^{\frac{W}{2}} \abs{\hat{g}_{tx,\thetav}(f)}^2 df
\end{equation}
where $W \coloneqq \frac{1}{T}$ is the bandwidth of the radio system, and $\hat{g}_{tx,\thetav}(f)$ the Fourier transform of $g_{tx,\thetav}(t)$.

Concerning the \gls{PED}, we denote by $p(t) \coloneqq \abs{x(t)}^2$ the signal power at time $t$.
Motivated by Chebyshev's inequality
\begin{equation}
	\Pr\LB \abs{p(t) - \EE \LSB p(t) \RSB} > a \RB \leq \frac{\text{Var}\LB p(t) \RB}{a^2}
\end{equation}
we choose to constrain the variance of the power $\text{Var}\LB p(t)\RB$ to enforce the \gls{PED} to have limited dispersion, and as a consequence to increase the power amplifier efficiency.
As the sequence of modulated symbols $\sv$ is \gls{iid}, one has
\begin{multline}
	\label{eq:var}
	\text{Var}\LB p(t) \RB = \LB \mu_4 - \mu_2^2 \RB \sum_l \abs{g_{tx,\thetav}(t - lT)}^4
	\\+ \mu_2^2 \sum_l \sum_{i \neq l} \abs{g_{tx,\thetav}(t - lT)}^2 \abs{g_{tx,\thetav}(t - iT)}^2
	\\+ \tilde{\mu}_2 \sum_l \sum_{i \neq l} \LB g_{tx,\thetav}(t - lT) \RB^2 \LB g_{tx,\thetav}^*(t - iT) \RB^2
\end{multline}
where
\begin{align}
	\mu_4 &= \EE_{c \sim \Uc\LB\Cc\RB}\LSB \abs{c}^4 \RSB\\
	\mu_2 &= \EE_{c \sim \Uc\LB\Cc\RB}\LSB \abs{c}^2 \RSB\\
	\tilde{\mu}_2 &= \EE_{c \sim \Uc\LB\Cc\RB}\LSB c^2 \RSB \EE_{c \sim \Uc\LB\Cc\RB}\LSB \LB c^* \RB^2 \RSB.
\end{align}
The variance of the \gls{PED} therefore depends on the constellation $\Cc$ and transmit filter $g_{tx,\thetav}(t)$.
Moreover, as a consequence of $\sv$ being \gls{iid}, $p(t)$ and $p(t + lT)$, $l \in \ZZ$, share the same probability distribution for any $t$ and $t + lT$ distant by at least $\frac{D}{2}$ from the signal temporal edges.
Hence, it is enough to focus on a single period $\LB -\frac{T}{2}, \frac{T}{2} \RB$.

Maximizing $R$ with constraints on the \gls{ACLR} and \gls{PED} leads to the optimization problem that we aim to solve, and which can be formally expressed as follows
\begin{subequations}
\begin{align}
& \underset{\Cc,\thetav,\psiv,\gammav}{\text{maximize}}
& & R(\Cc,\thetav,\psiv,\gammav) \label{eq:prob} \tag{P}\\
& \text{subject to}
& &\int_{-\frac{D}{2}}^{\frac{D}{2}} \abs{g_{tx,\thetav}(t)}^2 dt = 1 \label{eq:pc1} \tag{Pa}\\
& & & \EE_{c \sim \Uc\LB\Cc\RB} \LSB \abs{c}^2 \RSB = 1 \label{eq:pc2} \tag{Pb}\\
& & & \text{ACLR}(\thetav) = \epsilon_A \label{eq:pc3} \tag{Pc}\\
& & & V(\thetav, \Cc) = \epsilon_V \label{eq:pc4} \tag{Pd}
\end{align}
\end{subequations}
where~\eqref{eq:pc1} and~\eqref{eq:pc2} constrain the energy of the waveform,~\eqref{eq:pc3} enforces the \gls{ACLR} to equal a predefined value $\epsilon_A$, and~\eqref{eq:pc4} enforces the average \gls{PED} variance
\begin{equation}
	V(\thetav, \Cc) \coloneqq \EE_{t\sim\Uc\LB-\frac{T}{2},\frac{T}{2}\RB} \LSB \text{Var}\LB p(t) \RB \RSB
\end{equation}
to equal a predefined value $\epsilon_V$.

\section{End-to-end waveform learning}
\label{sec:e2e}

The key idea of end-to-end learning of communication systems~\cite{8054694} is to implement a transmitter, channel, and receiver as a single \gls{NN}, and to jointly optimize the trainable parameters of the transmitter and receiver for a specific channel model.

\subsection{Trainable transmit and receive filters}

Solving~\eqref{eq:prob} requires being able to evaluate with low-complexity and for any values of $\thetav$ and $\psiv$ the function $\alpha(t)$~\eqref{eq:alpha} and the noise covariance~\eqref{eq:noise_cov} to implement the channel transfer function~\eqref{eq:ch}.
It also requires having low-complexity implementations of the in-band energy function~\eqref{eq:inb}, the variance of the \gls{PED}~\eqref{eq:var}, and to ensure the transmit filter and constellation have unit energy~\eqref{eq:pc1}-\eqref{eq:pc2}.

As computing some of these quantities requires integration, using \glspl{NN} to implement the transmit and receive filters leads to intractable calculations.
One option would be to approximate these integrals using, e.g., Monte Carlo estimation or Riemann sums, but these approaches are computationally demanding and only result in approximations.

To accurately and efficiently implement trainable transmit and receive filters, we use the functions $\LP \text{sinc} \LB Df - s \RB \RP_{s \in \ZZ}$, which form a basis in the frequency domain for functions time-limited to $\LB -\frac{D}{2}, \frac{D}{2} \RB$.
Therefore, the trainable transmit and receive filters are defined in the frequency domain as
\begin{align}
	\hat{g}_{tx,\thetav}(f) &\coloneqq \sqrt{C(\thetav)} \sum_{s=-S}^{S} \theta_s \text{sinc} \LB Df - s \RB \label{eq:ftx}\\
	\hat{g}_{rx,\psiv}(f) &\coloneqq \sum_{s=-S}^{S} \psi_s \text{sinc} \LB Df - s \RB \label{eq:frx}
\end{align}
where $S$ controls the number of trainable parameters of the transmit and receive filters.
Note that higher values for $S$ lead to more degrees of freedom for the trainable filters, at the cost of a higher complexity at training.
Moreover, the transmit and receive filters are not required to have the same number of parameters $2S+1$ or the same duration $D$, as it was done in this work for convenience.
In~\eqref{eq:ftx}, $C(\thetav)$ is a normalization constant that ensures the transmit filter has unit energy~\eqref{eq:pc1}, for which an expression is given later in this section.
Taking the inverse Fourier transform of~\eqref{eq:ftx} and~\eqref{eq:frx} leads to the time-domain expressions of the trainable filters
\begin{align}
	g_{tx,\thetav}(t) &= \frac{\sqrt{C(\thetav)}}{D} \text{rect}\LB \frac{t}{D} \RB \sum_{s=-S}^{S} \theta_s e^{j2\pi\frac{s}{D}t} \label{eq:tx}\\
	g_{rx,\psiv}(t) &= \frac{1}{D}\text{rect}\LB \frac{t}{D} \RB \sum_{s=-S}^{S} \psi_s e^{j2\pi\frac{s}{D}t}. \label{eq:rx}
\end{align}

A benefit of using this approach to implement the transmit filter is that no additional complexity is required on the transmitter side.
Moreover, all the quantities required for training can be exactly obtained through direct calculations:
\begin{align}
	&C(\thetav) = \frac{D}{\thetav\htp\thetav} \label{eq:c_norm}\\
	&\alpha(t) = \begin{cases}
	\frac{\sqrt{C(\thetav)}}{D} \thetav\tp\Am(t)\psiv & \text{ if } t \in \LB -D, D \RB\\
	0 & \text{ otherwise}\\
	\end{cases}\\
	&{\small\EE \LSB n_m n_{m+l}^* \RSB = \begin{cases}
	\frac{N_0}{D} \psiv\tp \Am'\LB lT \RB \psiv^* & \text{ if } lT \in \LB -D, D \RB\\
	0 & \text{ otherwise}\\
	\end{cases}}\\ 
	&E_I(\thetav) = C(\thetav) \thetav\htp \Em \thetav
\end{align}
where $\Am(t)$ is a $(2S+1) \times (2S+1)$ complex-valued matrix whose coefficients are given by
\begin{multline}
	A(t)_{s_1,s_2} = \\
	\begin{cases}
	e^{j2\pi\frac{s_1}{D}t}\Delta(t) \text{ if } s_1 = s_2\\
	e^{j\pi\LB 2\frac{s_1}{D}t + (s_2-s_1)\Sc(t)\RB}\frac{\sin \LB \pi(s_2-s_1)\Delta(t) \RB}{\pi(s_2 - s_1)} \text{ otherwise}
	\end{cases}
\end{multline}
where $-S \leq s_1,s_2 \leq S$, $\Delta(t) = L_{max}(t) - L_{min}(t)$, and $\Sc(t) = L_{max}(t) + L_{min}(t)$, with $L_{max}(t) = \min\LP \frac{1}{2} ; \frac{t}{D} + \frac{1}{2} \RP$, and $L_{min}(t) = \max\LP -\frac{1}{2} ; \frac{t}{D} - \frac{1}{2} \RP$.
Similarly, $\Am'(t)$ is a $(2S+1) \times (2S+1)$ complex-valued matrix whose coefficients are given by
\begin{multline}
	A'(t)_{s_1,s_2} = \\
	\begin{cases}
	e^{j2\pi\frac{s_1}{D}t}\Delta'(t) \text{ if } s_1 = s_2\\
	e^{j\pi\LB 2\frac{s_1}{D}t + (s_1-s_2)\Sc'(t)\RB}\frac{\sin \LB \pi(s_1-s_2)\Delta'(t) \RB}{\pi(s_1 - s_2)} \text{ otherwise}
	\end{cases}
\end{multline}
where $-S \leq s_1,s_2 \leq S$, $\Delta'(t) = L_{max}'(t) - L_{min}'(t)$, and $\Sc'(t) = L_{max}'(t) + L_{min}'(t)$, with $L_{max}'(t) = \min\LP \frac{1}{2} ; -\frac{t}{D} + \frac{1}{2} \RP$, and $L_{min}'(t) = \max\LP -\frac{1}{2} ; -\frac{t}{D} - \frac{1}{2} \RP$.
Finally, $\Em$ is a $(2S+1) \times (2S+1)$ real-valued matrix whose coefficients are given by
\begin{equation}
	E_{s_1,s_2} = \int_{-\frac{W}{2}}^{\frac{W}{2}} \text{sinc}\LB Df - s_1 \RB \text{sinc}\LB Df - s_2 \RB
\end{equation}
and which can be pre-computed prior to training.

\subsection{Trainable constellation}

The trainable constellation consists of a set of $2^K$ complex numbers denoted by $\tilde{\Cc}$, and corresponding to the constellation points.
To transmit the data, $\tilde{\Cc}$ is normalized to ensure that the constraint~\eqref{eq:pc2} is satisfied, i.e.,
\begin{equation}
	\label{eq:const_norm}
	\Cc = \frac{\tilde{\Cc}}{\sqrt{2^{-K}\sum_{c\in\tilde{\Cc}}\abs{c}^2}}.
\end{equation}

\subsection{Neural network receiver}

\begin{figure}
     \begin{subfigure}[b]{\columnwidth}
         \centering
		\includegraphics[width=1.0\textwidth]{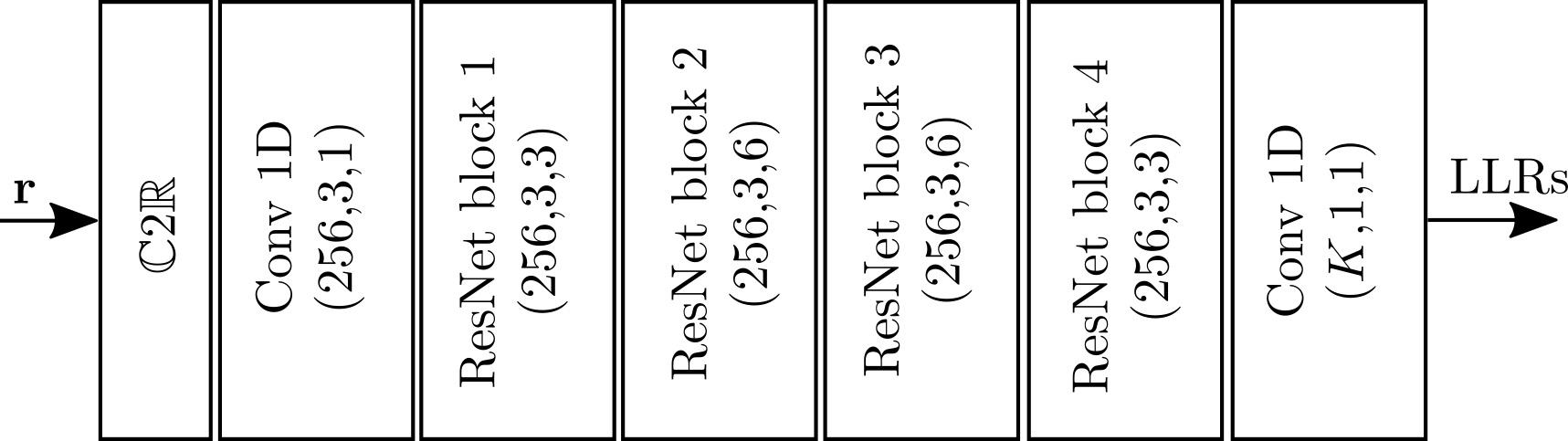}
         \caption{Architecture of the \gls{NN}-based detector. Below each layer name figures the (number of kernels, kernel size, dilatation rate).}
     \end{subfigure}

     \begin{subfigure}[b]{\columnwidth}
         \centering
         \includegraphics[width=0.75\textwidth]{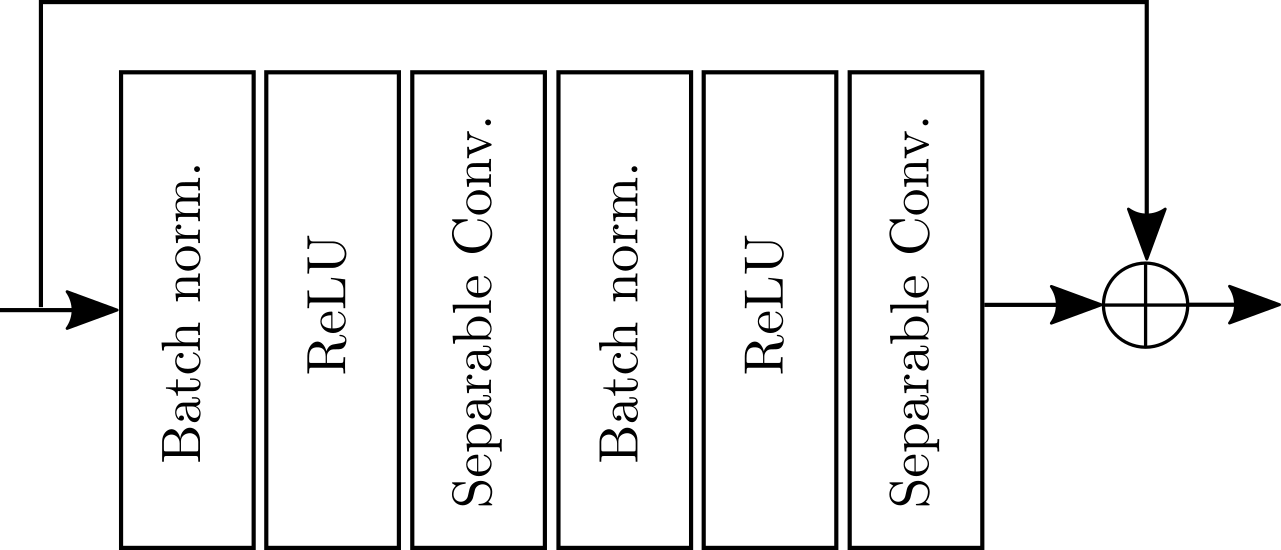}
         \caption{Architecture of a ResNet block.}
     \end{subfigure}
        \caption{The receiver is implemented by a convolutional \gls{NN}.}
        \label{fig:nnrx}	
\end{figure}

Fig.~\ref{fig:nnrx} shows the architecture of the \gls{NN} that implements the detector in the trainable end-to-end system.
It is a residual 1-dimensional convolutional \gls{NN}~\cite{He_2016_CVPR}, that takes $\rv \in \CC^{N}$ as input, and outputs a 2-dimensional tensor of \glspl{LLR} with shape $N \times K$.
The first layer converts the complex-valued input of length $N$ to a real-valued tensor with shape $N \times 2$, by stacking the real and imaginary parts.
Separable convolutional layers are leveraged to reduce the complexity of the \gls{NN}, while maintaining its performance.
Zero-padding is used to ensure the output has the same length as the input, and dilatation to increase the receptive field of the convolutional layers.

\subsection{Training algorithm}

Training of the end-to-end system requires evaluation of the achievable rate $R$~\eqref{eq:R}.
As usually done in end-to-end learning (e.g.,~\cite{aoudia2021trimming}), maximization of $R$ is achieved through the minimization of the total binary-cross-entropy
\begin{multline}
	\label{eq:loss}
	\Lc(\Cc,\thetav,\psiv,\gammav)\\
	\coloneqq -\frac{1}{N} \sum_{m=0}^{N-1} \sum_{k=0}^{K-1} \EE \LSB \log{Q_{\gammav}\LB B_{m,k} | \rv \RB} | \Cc,\thetav,\psiv \RSB
\end{multline}
which is related to $R$ by
\begin{equation}
	\Lc(\Cc, \thetav,\psiv,\gammav) = K - R(\Cc,\thetav,\psiv,\gammav).
\end{equation}
Since~\eqref{eq:loss} is numerically difficult to compute, it is estimated through Monte Carlo sampling by
\begin{multline}
	\Lc(\Cc,\thetav,\psiv,\gammav)\\
	\approx -\frac{1}{MN} \sum_{i=0}^{M-1} \sum_{m=0}^{N-1} \sum_{k=0}^{K-1} \log{Q_{\gammav}\LB B_{m,k}^{[i]} \lvert \rv^{[i]} \RB}
\end{multline}
where $M$ is the batch size, i.e., the number of samples used to compute the estimate of $\Lc$, and the superscript $[i]$ is used to refer to the $i^{th}$ example within a batch.

\begin{algorithm}
\caption{Training algorithm}
\label{alg:alt}
\begin{algorithmic}[1]
\State Initialize $\tilde{\Cc}$, $\thetav$, $\psiv$, $\gammav$, $\eta^{[0]}$, $\lambda_A^{[0]}$, and $\lambda_V^{[0]}$.
  \For{$u=0,\cdots$}
    \State Perform SGD on $\Lc_A\LB \Cc,\thetav,\psiv,\gammav;\lambda_A^{[u]},\lambda_V^{[u]}, \eta^{[u]} \RB$\label{lst:sgd}
    \State \LineComment{Update Lagrange multipliers:}
    \State $\lambda_A^{[u+1]} \gets \lambda_A^{[u]} - \eta^{[u]} \LB \text{ACLR}(\thetav) - \epsilon_A \RB$
    \State $\lambda_V^{[u+1]} \gets \lambda_V^{[u]} - \eta^{[u]} \LB V(\thetav,\Cc) - \epsilon_V \RB$
    \State \LineComment{Update penalty parameter:}
    \State Set $\eta^{[u+1]}$ such that $\eta^{[u+1]} > \eta^{[u]}$ \label{lst:eta}
  \EndFor
\end{algorithmic}
\end{algorithm}

\begin{figure*}
	\centering	
     \begin{subfigure}[b]{0.47\linewidth}
         \centering
	\includegraphics[width=1.0\textwidth]{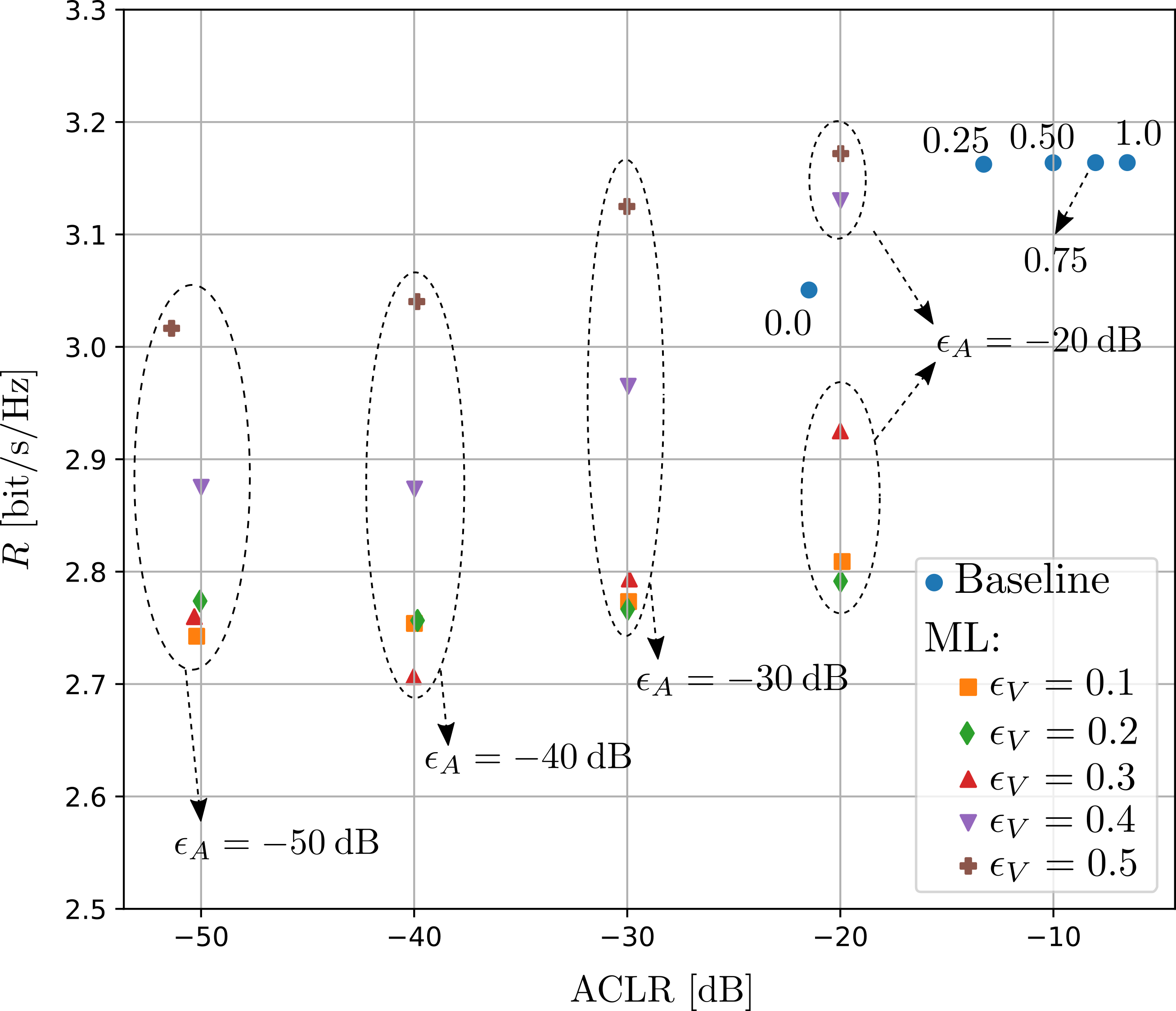}
	\caption{\glspl{ACLR} and rates achieved by the compared schemes.The roll-off factor value is indicated near each point for the baseline.\label{fig:results_r_aclr}}
     \end{subfigure}
     \quad\quad
     \begin{subfigure}[b]{0.48\linewidth}
         \centering
         \includegraphics[width=1.0\textwidth]{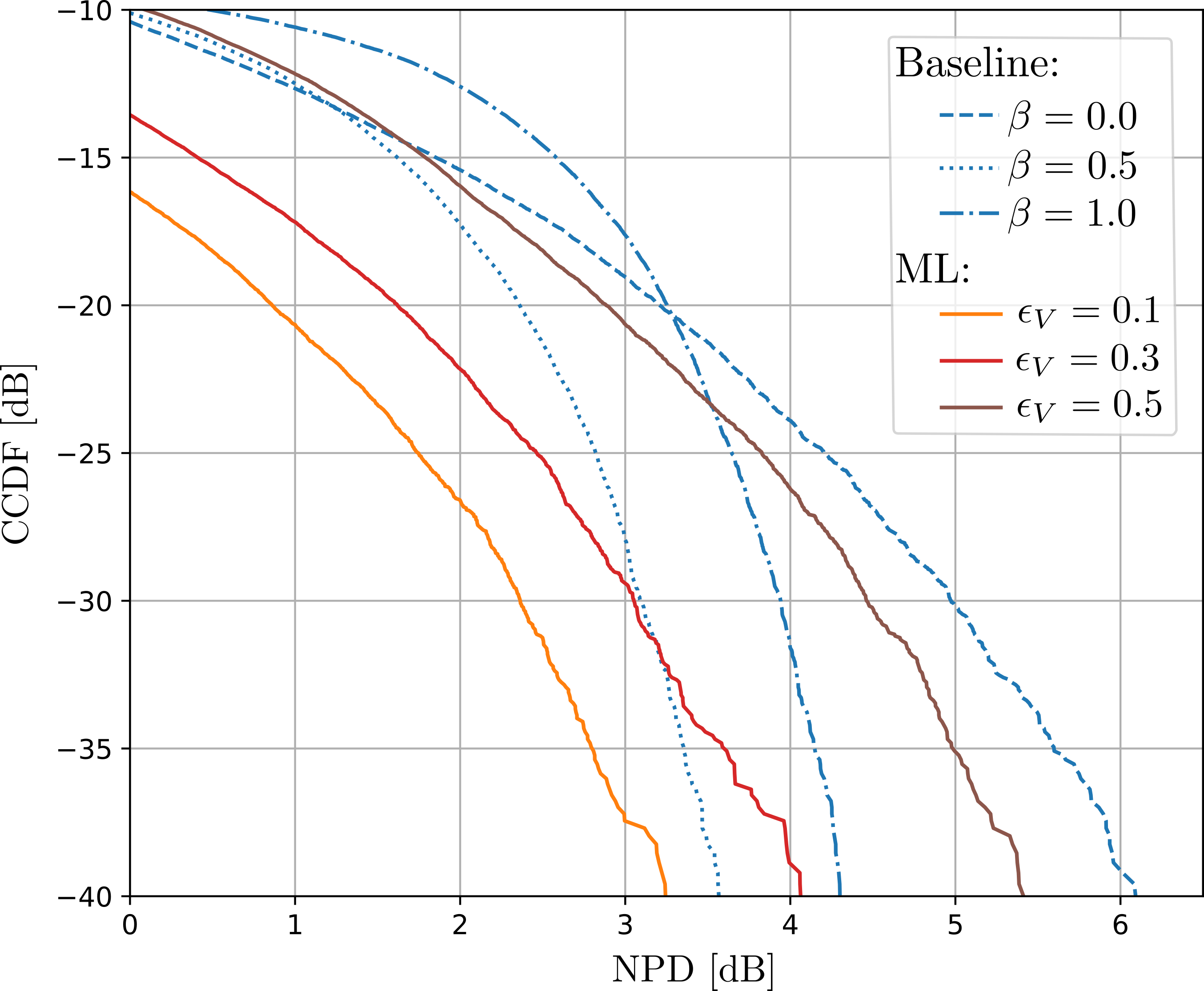}
         \caption{CCDF of the NPD of the baseline and ML schemes for $\epsilon_A = -30\:$dB.\label{fig:results_pow}}
     \end{subfigure}
     \caption{Rate, ACLR, and NPD achieved by the baseline and ML schemes.\label{fig:res}}
\end{figure*}

Finding a local optimum of~\eqref{eq:prob} is made challenging by the constraints~\eqref{eq:pc1}--\eqref{eq:pc4}.
The constraint~\eqref{eq:pc1} is enforced through the normalization constant $C(\thetav)$~\eqref{eq:c_norm} in~\eqref{eq:tx}, and the constraint~\eqref{eq:pc2} is enforced through the normalization of the constellation~\eqref{eq:const_norm}.
To handle the constraints on the \gls{ACLR}~\eqref{eq:pc3} and \gls{PED}~\eqref{eq:pc4}, the augmented Lagrangian method~\cite[Chapter~17]{nocedal2006numerical} is used, which, when applied to our setup, is shown in Algorithm~\ref{alg:alt}.
The augmented Lagrangian method consists in solving a sequence of unconstrained optimization problems (line~\ref{lst:sgd}), each aiming at minimizing the augmented Lagrangian
\begin{multline}
	\Lc_A \LB \Cc,\thetav,\psiv,\gammav;\lambda_A^{[u]},\lambda_V^{[u]}, \eta^{[u]} \RB \coloneqq \Lc(\Cc,\thetav,\psiv,\gammav)
	\\- \lambda_V^{[u]}\LB V(\thetav,\Cc) - \epsilon_V \RB - \lambda_A^{[u]}\LB \text{ACLR}(\thetav) - \epsilon_A \RB\\
	+ \frac{\eta}{2}\LB \LB V(\thetav,\Cc) - \epsilon_V \RB^2 + \LB \text{ACLR}(\thetav) - \epsilon_A \RB^2 \RB
\end{multline} 
where the superscript $[u]$ refers to the $u^{th}$ iteration, $\lambda_A$ and $\lambda_V$ are the Lagrange multipliers, and $\eta$ a positive penalty parameter that is progressively increased (line~\ref{lst:eta}).
At each iteration, minimizing the augmented Lagrangian is approximately achieved through \gls{SGD}.

\section{Evaluation of end-to-end learning}
\label{sec:results}

\begin{figure*}
	\centering	
     \begin{subfigure}[b]{0.56\linewidth}
         \centering
	\includegraphics[width=1.0\textwidth]{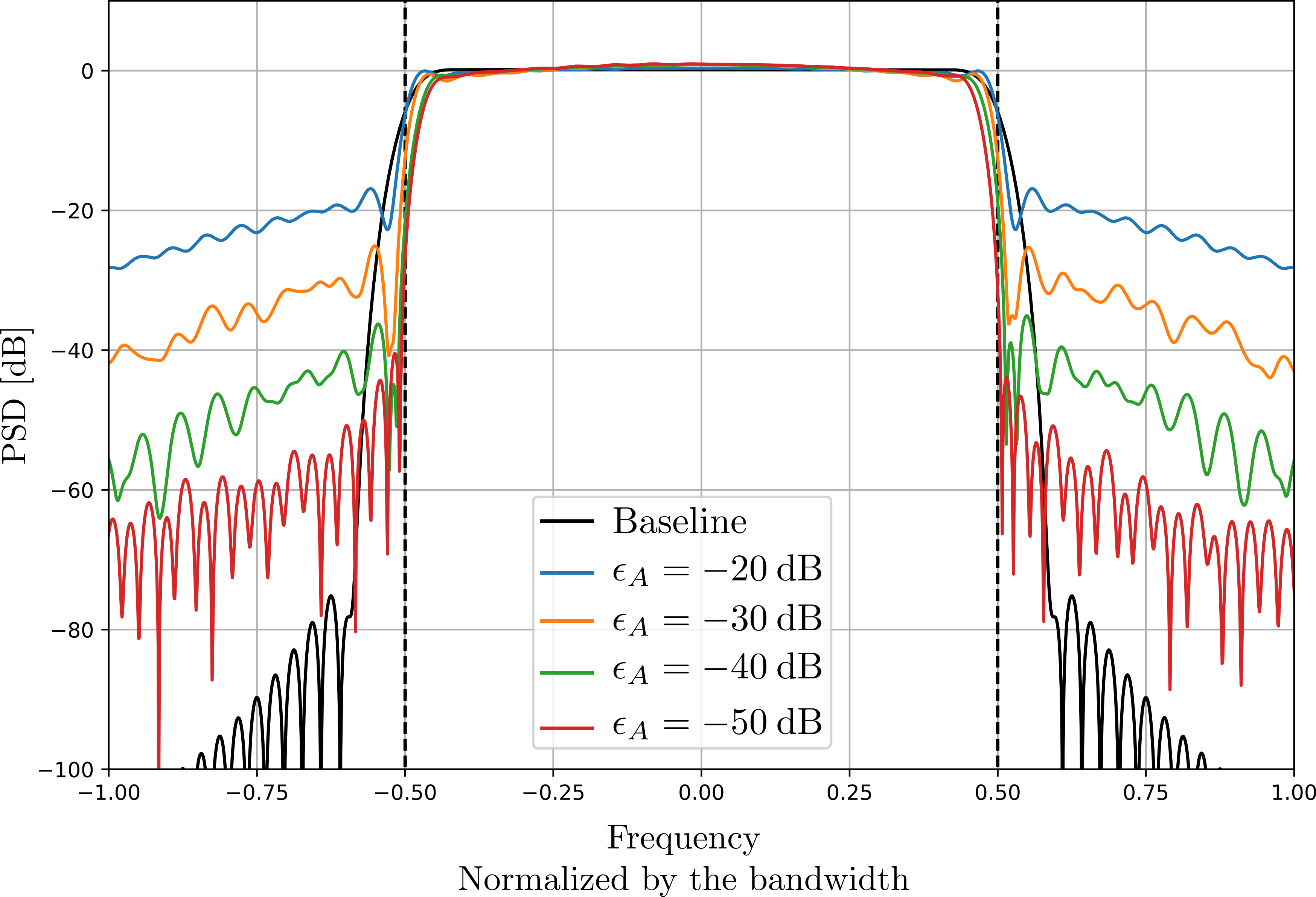}
	\caption{PSD of the baseline and learned filters ($\epsilon_V = 0.5$).\label{fig:spec}}
     \end{subfigure}
     \quad\quad
     \begin{subfigure}[b]{0.39\linewidth}
         \centering
         \includegraphics[width=1.0\textwidth]{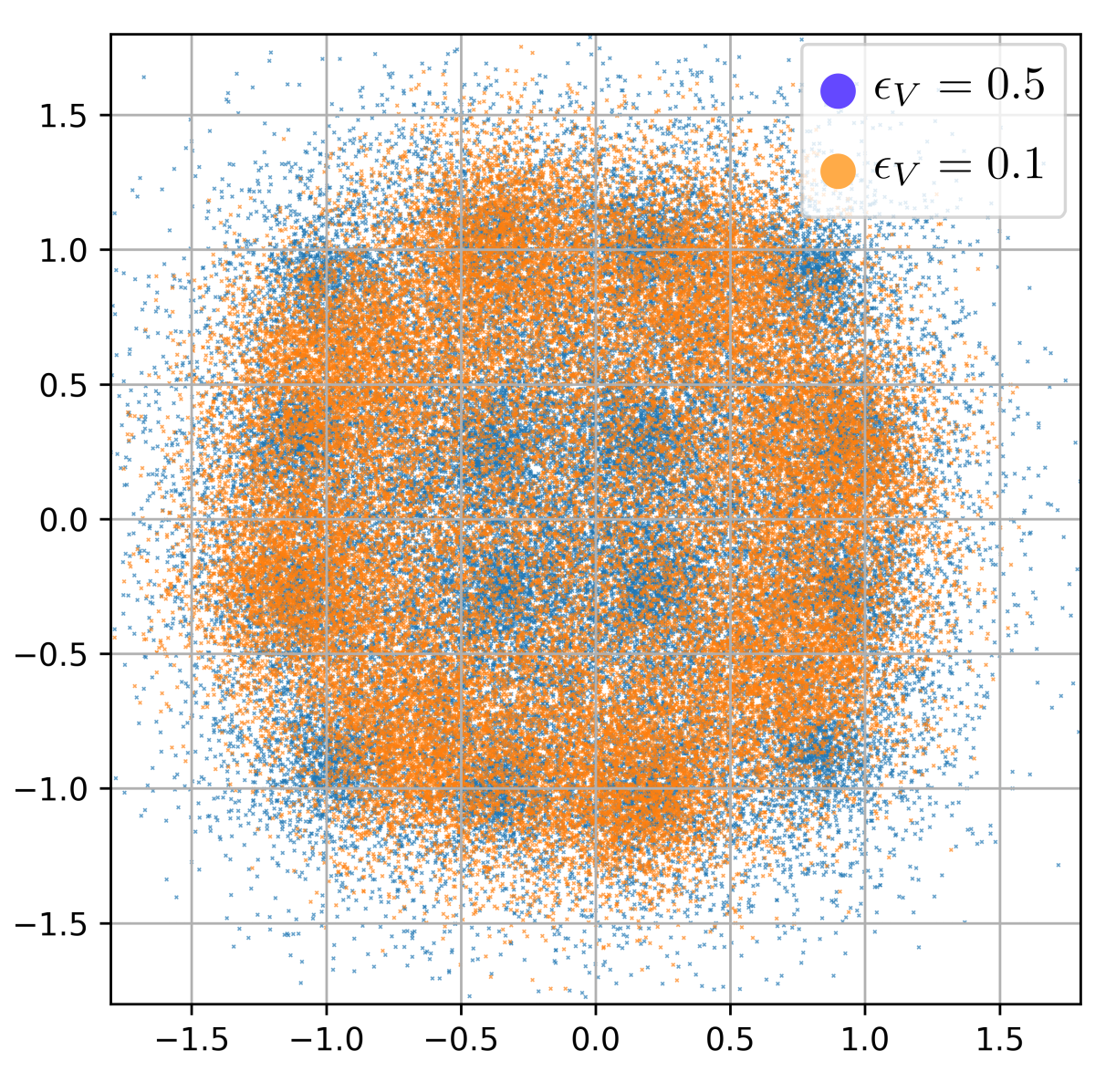}
         \caption{Transmitted signal $x(t)$ ($\epsilon_A = -30\:$dB).\label{fig:const}}
     \end{subfigure}
     \caption{Visualization of the transmitted signal with the learned waveform.\label{fig:viz}}
\end{figure*}

We will now present the results of the simulations we have conducted to evaluate the end-to-end learning scheme introduced in the previous section, referred to as the \gls{ML} scheme.
This section first provides details on the training and evaluation setup.
The ML scheme is then compared to the baseline.

\subsection{Evaluation setup}

An \gls{AWGN} channel implemented as described in Section~\ref{sec:sys_mod} was considered to train and to benchmark the ML scheme against the baseline presented in Section~\ref{sec:bsl}.
Note that the approach proposed in Section~\ref{sec:e2e} can be applied to any channel model whose output is differentiable with respect to its input, such as multi-paths channels.
However, evaluation on such channel models is left for future investigation.
The carrier frequency was set to $3.5\:$GHz, and the bandwidth to $W = 5\:$MHz.
The duration of the transmit and receive filters were set to $D = 32T$, and the number of parameters was set to $S = 100$ for the trainable filters.
The block length was set to $N = 990$ symbols, and the modulation order to $16$ ($K = 4$).
A Gray-labeled 16\gls{QAM} was used for the baseline.
The \gls{NN}-based receiver operates on the entire block of $N$ symbols.
The \gls{SNR} is defined by $\text{SNR} \coloneqq \frac{1}{N_0}$, and was set to $10\:$dB.
Training of the ML system was carried out using the Adam~\cite{adam} optimizer to perform \gls{SGD}, and with batches of size $M = 10$ and a learning rate set to $10^{-3}$.
The penalty parameter was initialized to $\eta^{[0]} = 10^{-2}$, and increased following a multiplicative schedule $\eta^{[u+1]} = 1.2\eta^{[u]}$.
The Lagrange multipliers were both initialized to $0$.

\subsection{Evaluation results}

Fig.~\ref{fig:results_r_aclr} shows the rates $R$ and \glspl{ACLR} achieved by the baseline and the ML schemes, for different values of $\epsilon_A$ and $\epsilon_V$.
Concerning the baseline, setting the roll-off factor $\beta$ to 0 gives the lowest \gls{ACLR}, at the cost of a rate loss due to the \gls{ISI} introduced by the windowing.
Higher values of $\beta$ lead to lower ripples making the filter more robust to windowing, at the cost of higher \glspl{ACLR}.
The \gls{ML} approach, on the other hand, allows accurate control of the \gls{ACLR} through the parameter $\epsilon_A$.
Moreover, it enables \glspl{ACLR} up to $30\:$dB lower than the ones of the baseline, without significant loss of rate assuming high enough values of $\epsilon_V$.

The constraint on the \gls{PED}~\eqref{eq:pc4} has the strongest impact of the rate $R$.
To quantify the impact of this constraint, Fig.~\ref{fig:results_pow} shows the \gls{CCDF} of the \gls{NPD},
\begin{equation}
	\text{NPD} \coloneqq \frac{\abs{p(t) - \EE\LSB p(t) \RSB}}{\EE\LSB p(t) \RSB}.
\end{equation}
As one can see, for the lowest values of $\epsilon_V$, the ML scheme achieves the lowest \gls{NPD}, at the cost of a rate loss (Fig.~\ref{fig:results_r_aclr}).
This leads to a \gls{PAPR} of $7.53\:$dB for the baseline when $\beta = 0$, whereas for $\epsilon_V = 0.4$ and $\epsilon_A = -30\:$dB, the \gls{ML} approach achieves a \gls{PAPR} of $6.45\:$dB, while enabling a significantly smaller \gls{ACLR} and a higher rate.

Fig.~\ref{fig:spec} shows the \gls{PSD} of the learned filters ($\epsilon_V = 0.5$) and of the baseline ($\beta = 0$).
As expected, the learned filter emits less in the adjacent bands, especially for low values of $\epsilon_A$, which translates into lower \glspl{ACLR}.
Concerning the constraint on the \gls{PED}, Fig.~\ref{fig:const} shows samples from the transmitted signal $x(t)$ for two different values of $\epsilon_V$.
As one can see, with low values of $\epsilon_V$, the samples are concentrated in a ``doughnut-like'' shape, corresponding to a constant envelope signal, and which would lead to a higher power amplifier efficiency.

\section{Conclusion}
\label{sec:conclu}

We have presented an end-to-end learning method for learning waveforms through joint optimization of constellation and pulse shaping to maximize an achievable information rate, with constraints on the \gls{ACLR} and power envelope.
Our simulations results show that the proposed approach enables lower \gls{ACLR} and less dispersive power envelopes while maintaining competitive rates.
Moreover, the end-to-end leaning method does not incur any additional complexity on the transmitter side, as it relies on conventional architectures, and can be applied to any channel model.
The presented results are therefore promising for evaluation on other channel models and to design waveforms for emerging beyond 5G services.

\bibliographystyle{IEEEtran}
\bibliography{IEEEabrv,bibliography}

\end{document}